\documentclass[
twocolumn,
]{ceurart}

\usepackage{multicol}
\usepackage{balance}
\usepackage{flushend}
\usepackage{minted}
\setminted{breaklines=true}

\begin{document}

\copyrightyear{2024}
\copyrightclause{Copyright for this paper by its authors.
  Use permitted under Creative Commons License Attribution 4.0
  International (CC BY 4.0).}

\conference{Joint Proceedings of IVA'24 Workshops and Doctoral Consortium, September 15--19, 2024, Glasgow, UK}

\title{Chatbots to strengthen democracy: An interdisciplinary seminar to train identifying argumentation techniques of science denial}

\author[1]{Ingo Siegert}[%
orcid=0000-0001-7447-7141,
email=siegert@ovgu.de,
]
\cormark[1]
\fnmark[1]
\address[1]{Mobile Dialog Systems,
  Otto-von-Guericke-University Magdeburg, Germany}
  
\author[2]{Jan Nehring}[
email=jan.nehring@dfki.de,
]
\fnmark[1]
\address[2]{German Research Centre for Artificial Intelligence, Berlin}

\author[3]{Aranxa Márquez Ampudia}[%
]
\fnmark[1]
\address[3]{Quality and Usability Lab, TU Berlin, Germany}

\author[1]{Matthias Busch}[
orcid=0000-0003-1420-3602,
email=matthias.busch@ovgu.de,
]
\fnmark[1]

\author[3]{Stefan Hillmann}[
email=stefan.hillmann@tu-berlin.de,
]
\fnmark[1]

\cortext[1]{Corresponding author.}
\fntext[1]{These authors contributed equally.}
\begin{abstract}
In recent times, discussions on social media platforms have increasingly come under scrutiny due to the proliferation of science denial and fake news. Traditional solutions, such as regulatory actions, have been implemented to mitigate the spread of misinformation. However, these measures alone are not sufficient. 
To complement these regulatory efforts, educational approaches are becoming essential in empowering users to critically engage with and counteract misinformation. Conversation training, through either serious games or personalized methods, has emerged as a promising strategy to help users identify and handle science denial and toxic conversation tactics. 
This paper suggests an interdisciplinary seminar to explore the suitability of Large Language Models (LLMs) acting as a persona of a science denier to develop Chatbots able to support people in identifying  misinformation and improving user resilience against toxic online interactions. 

In the seminar, groups of four to five students will develop an AI-based chatbot that enables realistic interactions with science-denial argumentation structures. The task involves planning the setting, integrating a Large Language Model to facilitate natural dialogues, implementing the chatbot using the RASA framework, and evaluating the outcomes in a user study. This requires not only learning and applying the AI-supported framework to ensure the chatbot understands inputs and responds appropriately to advance the interaction, but also developing the chatbot’s personality to effectively engage users in addressing science denial. It is crucial that users understand what they need to do during the interaction, how to conclude it, and how the relevant information is conveyed. Hereby the seminar does not aim to develop chatbots to practice debunking science denier claims in conversations, moreover, it serves as a seminar to teach newest AI-technology to students and to test the feasibility of this idea for future applications.
The chatbot seminar is conducted as a hybrid, parallel master’s module at the participating educational institutions.
\end{abstract}


\maketitle

\section{Introduction}

Many people in today's society are confronted with the challenge of seeing someone in their close circle—such as friends, neighbors, or family members—drift toward far-right political beliefs. This shift is often accompanied by a belief in conspiracy theories and the consumption of misinformation. These concerned individuals want to engage in discussions with those drifting into these views, but such conversations often escalate quickly, leading to stress in personal relationships and, ultimately, contributing to political polarization. Our goal is to develop chatbots that play the role of someone adopting far-right or conspiracy-theory beliefs, allowing users to practice and refine their conversation skills in these challenging situations.

Misinformation poses significant threats to various aspects of society. In the realm of public health, the spread of false information can lead to widespread health misconceptions, vaccine hesitancy, and poor health outcomes~\cite{info:doi/10.2196/17187}. For example, during the COVID-19 pandemic, misinformation about the virus and vaccines led to increased resistance to vaccination efforts, undermining public health initiatives~\cite{covid23}. Political stability is also jeopardized as misinformation can fuel polarization, erode trust in democratic institutions, and incite violence~\cite{doi:10.1177/20563051231177943}. False narratives and conspiracy theories often lead to political unrest and weaken the fabric of democratic societies~\cite{REfugeeGerman}. Furthermore, scientific literacy~\cite{WEI2023101910} suffers as misinformation distorts public understanding of scientific facts and undermines trust in scientific expertise, making it harder for the public to make informed decisions on critical issues such as climate change~\cite{FREILING2023107769}.

This phenomenon has highlighted the crucial responsibility of platform operators to implement appropriate measures and regulations. 
Platform operators bear a critical responsibility in curbing the spread of misinformation, fake news, and science denial. Governments are implementing several measures and regulations to address this issue~\cite{10.1093/oso/9780197692851.001.0001}. On the one hand, content moderation policies are being strengthened, employing both human reviewers and AI algorithms to identify and remove false information~\cite{doi:10.1177/2053951719897945}. On the other hand, volunteer fact-checker organizations help verify the accuracy of content, while flagging systems warn users about potentially misleading posts~\cite{doi:10.1177/00936502231206419,Fact-checkingPoland}. In addition, platform operators are enhancing transparency by providing more context on information sources and limiting the reach of accounts that repeatedly share falsehoods. But, as the previous mentioned impacts show, these actions are not enough, especially, as some social networks reduce or stop their efforts in this area or create their own networks~\cite{10.1093/hrlr/ngac009,Sawyer2018}.

On a governmental level, corresponding regulations are already being developed and enforced.  In the United States, the government has introduced legislation like the Honest Ads Act, which aims to increase transparency in online political advertisements~\cite{HonestAdsAct}. Additionally, agencies such as the Federal Trade Commission (FTC) actively work to combat deceptive practices online~\cite{FTCClimateChange}. In the United Kingdom, the Online Safety Bill seeks to hold social media companies accountable for harmful content, including misinformation, by imposing fines for non-compliance~\cite{bournemouth39084}. The UK government also collaborates with fact-checking organizations to monitor and address false information~\cite{doi:10.1177/14648849221078465}. The European Union has implemented the Digital Services Act, which mandates stricter content moderation and transparency requirements for online platforms~\cite{Tourkochoriti23}, and the Code of Practice on Disinformation, a voluntary agreement among major tech companies to curb the spread of false information~\cite{EUCodeofPractice}. These regulations are enforced through a combination of fines, mandatory reporting, and oversight by regulatory bodies. Despite these efforts, ongoing evaluation and adaptation of these regulations are necessary to keep pace with the rapidly evolving digital landscape. However, these measures do not immediately help individual users who encounter accounts that spread such content on social networks~\cite{Jansen_Kraemer_2023}. 
Furthermore, these governmental actions are also seen as a massive encroachment on civil rights in terms of surveillance techniques and censoring content~\cite{doi:10.1080/17577632.2022.2083870}. 

To counteract these issues on a personal level, various strategies can be employed. Training programs can be developed to enhance social network literacy, helping individuals recognize and critically evaluate false information~\cite{Livingstone2014,ncte:/content/journals/10.58680/rte201220670}. Commonly used are conversation training initiatives, such as serious games or personalized workshops, that can equip users with the skills to engage constructively with misinformation and handle toxic discussions. 
Previous studies have shown that individuals who undergo structured conversation training are better equipped to recognize and counteract false information~\cite{doi:10.1098/rsos.211719}. For instance, research by the University of Cambridge showed that participants who engaged in ``prebunking'' exercises, where they were exposed to common misinformation tactics, were more adept at identifying and rejecting fake news~\cite{doi:10.1177/20539517211013868}.  
Similarly, a study conducted by the RAND Corporation found that training individuals in techniques such as active listening and fact-based rebuttal significantly improved their ability to engage in constructive dialogue with those holding misinformation-based beliefs\footnote{\url{https://www.rand.org/research/projects/truth-decay/fighting-disinformation/search.html}}. 
In addition, serious games designed to simulate encounters with misinformation have been shown to enhance users' critical thinking skills and resilience to deceptive content~\cite{doi:10.1080/10494820.2023.2299999,GIANNAKOPOULOU2023FID}. These findings underscore the value of conversation training as a practical approach to reducing the spread and impact of misinformation in digital spaces. Furthermore, they are fostering a culture of continuous learning and critical skepticism can empower individuals to remain vigilant and informed in the face of evolving misinformation tactics. By integrating these approaches, individuals can become more resilient against the spread of misinformation and contribute to a more informed online community.

Simultaneously, the emergence of ChatGPT and other large language models (LLMs) has made it possible to conduct natural language dialogues with AI-based systems, simulating a variety of conversation types. In this context, a methodology known as prompt engineering~\cite{sahoo2024systematicsurveypromptengineering} has been developed, which enables the manipulation of LLMs to respond in specific ways and with particular characteristics~\cite{gu2023effectivenesscreatingconversationalagent}.

By addressing these aspects, this paper describes an interdisciplinary seminar to teach newest AI-technology to students and to test the potential of AI-based dialogue systems in combating the spread of misinformation and improving user resilience against toxic online interactions. 
Hereby, the seminar does not aim to develop chatbots to practice debunking science denier claims in conversations.

\section{Argumentation techniques of science denial}
Science denial refers to the rejection of scientific evidence and consensus, often in favor of personal beliefs, misinformation, or political ideology~\cite{Denialism}. It can manifest in various forms, from denying climate change and the efficacy of vaccines to rejecting evolutionary theory. Addressing science denial is not just a matter of presenting facts; it involves engaging in a constructive dialogue and employing effective argumentation strategies.

To better understand and address science denial, the concept of FLICC -- an acronym for Fake experts, Logical fallacies, Impossible expectations, Cherry picking, and Conspiracy theories -- was introduced~\cite{FLICC}. FLICC outlines a taxonomy of fallacies and argumentative reasoning used to undermine scientific understanding and spread doubt. Recognizing these tactics, which are not mutually exclusive, is crucial for anyone involved in science communication, as it helps to identify and counter denialist arguments effectively~\cite{Climate-debunking}.

\begin{figure}[h]
    \centering
    \includegraphics[width=0.95\columnwidth]{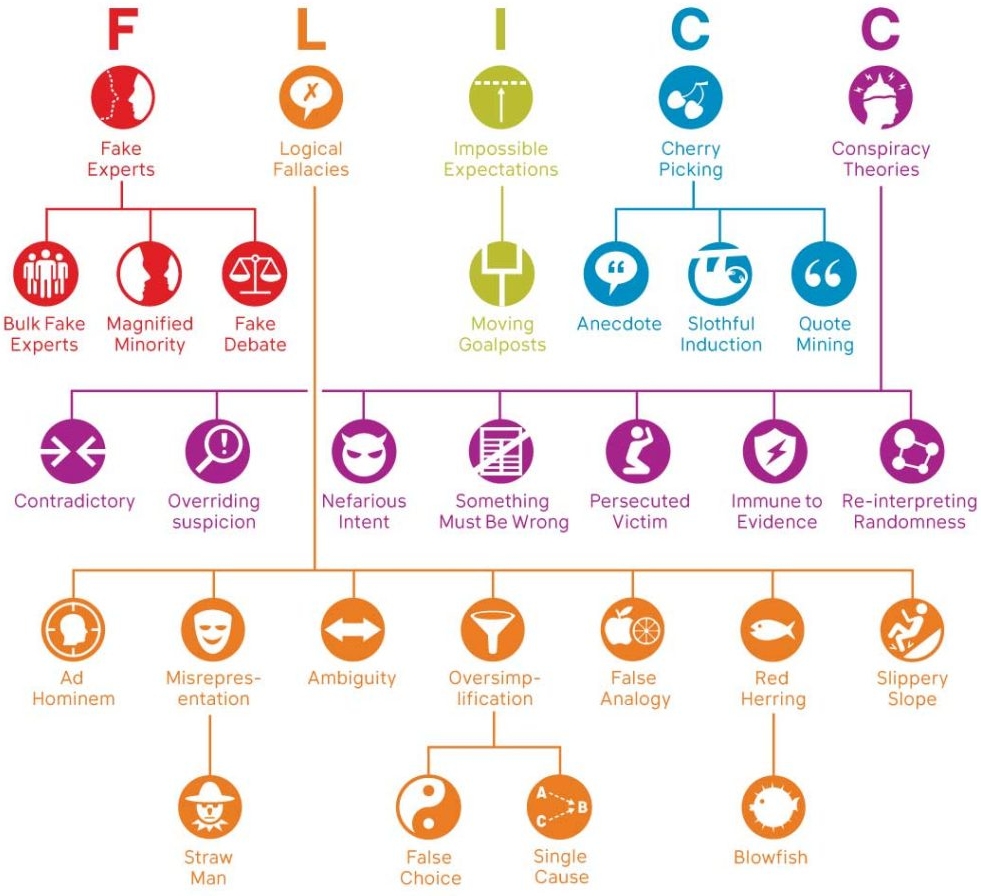}
    \caption{Overview of Techniques of Science Denial.}
    \label{fig:FLICCstrategies}
\end{figure}

Engaging in science denial requires a nuanced approach that combines a clear presentation of evidence with techniques to address misconceptions and biases. By understanding the strategies behind science denial and using tools like FLICC, communicators can foster more productive discussions and help bridge the gap between scientific consensus and public perception. Figure~\ref{fig:FLICCstrategies} depicts an overview of the FLICC strategies and the different sub-strategies.

\section{The Seminar ``Chatbots to Strengthen Democracy''}
The challenge now lies in enabling persons, confronted with discussion about topics dangerous for society and democracy, to recognize these argumentative strategies and developing appropriate strategies to deal with them. In this context, we believe that chatbot systems would be highly suitable for educating users about these strategies and learning how to manage this type of interaction. To take a first step in this direction, we have developed a seminar in which students from technical disciplines are tasked with creating a chatbot system for training in handling science denial strategies.

The seminar titled ``Chatbots to Strengthen Democracy'' took place in the summer term of 2024 and was a collaboration between the participating educational institutions and was the extension of a seminar that focused on designing chatbots for escape room-like games~\cite{Chatbotchallenge1}. 
The current seminar was designed for one semester and focuses on the practical application of AI methods combined with the development of project management skills, scientific evaluation, and presentation capabilities. This was achieved through a clearly defined objective. Students, working in groups of 4-5, were tasked with developing a chatbot system that employs conspiracy theory arguments on a specific topic and utilizes selected science denial strategies. The goal was to enable users to practice handling these strategies and ideally recognize them. In addition, users will be provided with an appropriate explanation and introduction to these strategies. 

The chatbot, with a specific background story, will use science denial argumentation techniques. Users must identify and counter these techniques during the conversation. The chatbot will then acknowledge the identified strategy and conclude the dialogue.

The seminar was divided into practical sessions, presentations, and self-study phases. Practical sessions cover the introduction of the Chatbot Challenge, instructors, and rules. Additional sessions included an introduction to the Chatbot-platform using RASA~\cite{10.1007/978-981-19-2130-8_69} and dynamic prompting with LLMs as answer generation tool, argumentation strategies, user study design, and frontend deployment. Clear rules ensured comparable starting conditions: all groups must use the provided docker system (comprising frontend, backend, and dialog manager), no images in responses, and the use only of the provided LLM. The evaluation was conducted in parallel with a predefined UX questionnaire, UEQ+~\cite{doi:10.1080/10447318.2023.2258026}. 
Presentations occurred concurrently at the participating educational institutions for the following milestones:
\begin{itemize}
    \item MS1: Present overall idea, first prototype.
    \item MS2: First complete chatbot version, results of user first study.
    \item MS3: Final user study evaluation, lessons learned.
\end{itemize}
    
At the end of the seminar each group had to write a final report summarizing the results of the project including their science denial topic, the identified argumentation techniques, technical implementations applied, user study design and outcome as well as lessons learned.

    
\section{Technical platform for the Seminar}

\begin{figure}[h]
    \centering
    \includegraphics[width=0.99\columnwidth]{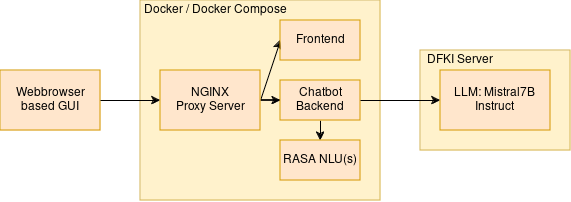}
    \caption{The chatbot challenge platform.}
    \label{fig:technicalplatform}
\end{figure}

As shown in Figure~\ref{fig:technicalplatform}, our technical platform consists of three basic components: First, a browser based graphical user interface (Frontend) in which users can interact with the various chatbots. Second, the chatbot backend, in which the students can implement their dialog logic and third, the Mistral 7B-Instruct \cite{jiang2023mistral7b} LLM.

The LLM has the ability to converse about almost any topic. Using prompting techniques \cite{schulhoff2024promptreportsystematicsurvey} one can create dialog systems from LLMs \cite{hudecek-dusek-2023-large}. We used a technique called “Dynamic Prompting” (paper currently under review) in which we composed prompts in the following form: First, the prompts contain the prompt template, which contains instructions regarding the chatbot personality and answering style. Then, the prompt contains the previous conversation in a theater script style prompt. The appendix (Section \ref{appendix:prompting:template}) contains an example of such a prompt.

We implemented the chat logic in the chatbot backend. Here, we used the RASA Natural Language Understanding system \cite{bunk2020dietlightweightlanguageunderstanding} for intent detection. Using RASA, the intent of a user can be classified (e.g., the user wants to tell the chatbot that he is wrong) as belonging to one of a list of predefined class labels. Based on these class labels, the prompt template will be changed. In this way, we implemented rules, such as: Normally, the chatbot shows behavior A. When the user contradicts the chatbot, the chatbot shows behavior B.

Some students used more advanced text analysis tools. They used the large language model and the few-shot learning prompting technique \cite{brown2020languagemodelsfewshotlearners} for analysis such as “Does the user contradict himself during the dialog” and incorporated the results in their dialog logic.

Using this technique, we enable the development of chatbots that can talk about almost every topic with a controllable answering style. With carefully engineered prompts, our chatbots can convincingly impersonate their roles.

The chatbot back-end can also send a signal to the user interface that the user has reached his goal. In this case, the user interface will, e.g., display a success message.

As LLM, we chose Mistral-7b-instruct  because this model contains almost no safety training \cite{interview_arthur_mensch}. Usually, LLMs such as ChatGPT contain safety training so they cannot generate fake news or other toxic content. Using Mistral-7b-instruct, we could generate such responses, which is needed for our approach of developing chatbots to train oneself in identifying science denial or fake news argumentation.

\section{Learning outcomes}
In the Challenge, students had the opportunity to engage with AI in a practical context. They were required to understand the underlying principles, such as training an NLU model and integrating the output of an LLM, while focusing on the results, which can be quickly validated through iterative chatbot versions. This approach emphasizes the conscious and proficient use of AI-supported dialogue system frameworks, increasingly essential in today's software landscape.

Additionally, students also gained critical skills, such as understanding and identifying the argumentation structure used, in this case, for science denial and reflecting on the risks and potential of using chatbots in combination with LLMs. 

The overall project framework imparted crucial skills and knowledge in UX design, project management, evaluation, and argumentation strategies. The competitive nature of the Challenge encouraged participants to compare different solutions, identify their advantages and disadvantages, and apply these insights to improve their own chatbots.


\section{Example bots}
When we engage in discussions, we sometimes find ourselves lacking the appropriate words, arguments, and even tone to explain why, for example, evolution has been scientifically proven and is not a lie. Additionally, we often fail to recognize where the other person’s arguments lose their logical foundation.

The latter topic was one example used by students during the challenge to develop the ``Evolution Denier'' chatbot. The creation of this chatbot system offered two learning processes: one for the students themselves and another for the users evaluating the chatbot.

The developing students first identified the most commonly used arguments to deny evolution. Then, with the help of the FLICC taxonomy, they categorized these arguments. When designing the instructions for interacting with the bot, they decided that users would be encouraged not only to explain why evolution is real, but also to read and understand the logical errors in the bot's arguments and point them out.

In the interaction with this bot, the users were challenged not only to think about facts, but also how to present them effectively. In this example, the developed chatbot system rewarded users for reacting civilly and for identifying the FLICC fallacies in the bot's responses. Hereby, students also implemented sentiment analyses on the users' input.

Another group developed a chatbot that mimics a climate change denier (ideally) changing his mind during the interaction. For that, the students introduced an internal belief level of the bot, quantizing the bot's belief in its own arguments.

Other chatbot topics were skepticism about green energy, vaccine hesitancy, incel community, the birds-aren't-real-conspiracy, intelligent design, and flat earth believer. This group also implemented a sentiment analysis using the  TextBlob library to assess the emotional tone of the user's input, allowing the chatbot to more accurately understand and respond to the user's feelings. The appendix provides images for example interactions with the last two chatbots mentioned. 

\section{Discussion and conclusion}
Born from the question of how technology can contribute to counteracting science denial and fake news, and how a platform can be created for self-training resilience, a seminar was developed. This seminar, conducted with students, aimed to explore whether and how such argumentation trainers could be implemented as chatbots using open source state-of-the-art LLMs.

The challenge began in April 2024 as an inter-university course. While students in their groups each develop a chatbot, we developed a portal into which the chatbots can be integrated. This allowed all participants to test their chatbots under the same conditions and to trial them in a protected environment. The challenge is limited in many ways to ensure that students focus on the aspects to be assessed and that the final evaluation of the results remains fair, as all use the same resources.

Differences emerged not only in the chosen science-denial topic but also in how the topic was addressed (explanation at the beginning, hints during the interaction, explanatory texts in the chatbot response) and in whether and how an assertive communication culture is maintained (sentiment analysis, point system for polite responses, etc.).
However, the first challenges in development are already becoming apparent. The trial-and-error approach, which is otherwise a viable method for small demonstrators, quickly reaches its time limits here.

Additionally, the following research questions arise for us, which we aim to answer after the seminar:

1)~Can large LLM powered chatbots mimic the persona of a citizen in a radicalization process? 
2)~Can we implement chatbots that train users in certain conversational strategies to counter the radicalization of peers? 
3)~How can this be implemented as a seminar for students of higher education? 
4)~What are the characteristics of this interaction? Specifically, we want to create an error taxonomy of our approach to explore the limitations of our technology.

The technology presented here can be adapted beyond the current seminar topic of science denial to address other issues such as the prevention of stereotypes and discrimination. This would allow students to recognize their own and others' biases, to stop perpetuation of negative stereotypes and discriminatory behavior towards minority groups without exposing themselves or others to these dialogues in real-world scenarios.



\bibliography{sample-ceur}

@article{doi:10.1080/10447318.2023.2258026,
author = {Ehsan Mortazavi, Philippe Doyon-Poulin, Daniel Imbeau and Jean-Marc Robert},
title = {Development and Validation of Four Social Scales for the UX Evaluation of Interactive Products},
journal = {International Journal of Human–Computer Interaction},
volume = {0},
number = {0},
pages = {1--14},
year = {2023},
publisher = {Taylor \& Francis},
doi = {10.1080/10447318.2023.2258026},
}

@InProceedings{10.1007/978-981-19-2130-8_69,
author="Kumari, Vijay
and Gosavi, Chinmay
and Sharma, Yashvardhan
and Goel, Lavika",
editor="Sharma, Harish
and Shrivastava, Vivek
and Kumari Bharti, Kusum
and Wang, Lipo",
title="Domain-Specific Chatbot Development Using the Deep Learning-Based RASA Framework",
booktitle="Communication and Intelligent Systems ",
year="2022",
publisher="Springer Nature Singapore",
address="Singapore",
pages="883--896",
isbn="978-981-19-2130-8"
}

@misc{FLICC,
    year= {2024},
    title = {A history of FLICC: the 5 techniques of science denial
},
    howpublished = {\url{https://skepticalscience.com/history-FLICC-5-techniques-science-denial.html}},
    note = {Accessed: 2024-06-01}
}

@book{Climate-debunking,
    author = {Margaret Orr and John Cook and Amanda Borth},
    title = {Climate Myth debunking. For Broadcast Meteorologists},
    publisher = {Center for Climate Change Communication} ,
    year = {2023} 
}

@article{Denialism,
    author = {Mark Hoofnagle and Chris Jay Hoofnagle},
    title = {What is Denialism?},
    journal = {SSRN},
    year = {2007},
    doi = {10.2139/ssrn.4002823},

}

@book{10.1093/oso/9780197692851.001.0001,
    author = {Gorwa, Robert},
    title = "{The Politics of Platform Regulation: How Governments Shape Online Content Moderation}",
    publisher = {Oxford University Press},
    year = {2024},
    month = {08},
    isbn = {9780197692851},
    doi = {10.1093/oso/9780197692851.001.0001},
}

@misc{sahoo2024systematicsurveypromptengineering,
      title={A Systematic Survey of Prompt Engineering in Large Language Models: Techniques and Applications}, 
      author={Pranab Sahoo and Ayush Kumar Singh and Sriparna Saha and Vinija Jain and Samrat Mondal and Aman Chadha},
      year={2024},
      eprint={2402.07927},
      archivePrefix={arXiv},
      primaryClass={cs.AI},
      url={https://arxiv.org/abs/2402.07927}, 
}

@misc{gu2023effectivenesscreatingconversationalagent,
      title={On the Effectiveness of Creating Conversational Agent Personalities Through Prompting}, 
      author={Heng Gu and Chadha Degachi and Uğur Genç and Senthil Chandrasegaran and Himanshu Verma},
      year={2023},
      eprint={2310.11182},
      archivePrefix={arXiv},
      primaryClass={cs.HC},
      url={https://arxiv.org/abs/2310.11182}, 
}

@InProceedings{GIANNAKOPOULOU2023FID,
author    = {Giannakopoulou, A. and Bevilacqua, L. and Gennarelli, A. and Kokole, L. and Lihtenvalner, K. and Walczak, I. and Vukcevic, R.},
title     = {FI.DO: FIGHTING FAKE NEWS AND DISINFORMATION- A SERIOUS GAME AND NEW METHODOLOGIES FOR TRAINING SENIOR CITIZENS},
series    = {17th International Technology, Education and Development Conference},
booktitle = {INTED2023 Proceedings},
isbn      = {978-84-09-49026-4},
issn      = {2340-1079},
doi       = {10.21125/inted.2023.0864},
url       = {https://doi.org/10.21125/inted.2023.0864},
publisher = {IATED},
location  = {Valencia, Spain},
month     = {6-8 March, 2023},
year      = {2023},
pages     = {3125-3132}}

@article{doi:10.1080/10494820.2023.2299999,
author = {Kristian Kiili, Juho Siuko and Manuel Ninaus},
title = {Tackling misinformation with games: a systematic literature review},
journal = {Interactive Learning Environments},
volume = {0},
number = {0},
pages = {1--16},
year = {2024},
publisher = {Routledge},
doi = {10.1080/10494820.2023.2299999},
}

@book{EUCodeofPractice,
    author = {James Pamment},
    title = {The EU Code of Practice
 on Disinformation:
 Briefing Note for the
 New EU Commission},
    publisher = {Carnegie Endowment for International Peace},
    year = {2020}
}

@article{Tourkochoriti23,
author = {Ioanna Tourkochoriti},
title = {The Digital Services Act and the EU as the Global Regulator of the Internet},
journal = {Chi. J. Int'l L.},
volume = {24},
number = {1},
pages = {129--147},
year = {2023},
url = {https://chicagounbound.uchicago.edu/cjil/vol24/iss1/7?utm_source=chicagounbound.uchicago.edu/cjil/vol24/iss1/7},
}

@article{doi:10.1177/20539517211013868,
author = {Melisa Basol and Jon Roozenbeek and Manon Berriche and Fatih Uenal and William P. McClanahan and Sander van der Linden},
title ={Towards psychological herd immunity: Cross-cultural evidence for two prebunking interventions against COVID-19 misinformation},

journal = {Big Data \& Society},
volume = {8},
number = {1},
pages = {20539517211013868},
year = {2021},
doi = {10.1177/20539517211013868},
}

@article{doi:10.1098/rsos.211719,
author = {Roozenbeek, Jon  and Traberg, Cecilie S.  and van der Linden, Sander },
title = {Technique-based inoculation against real-world misinformation},
journal = {Royal Society Open Science},
volume = {9},
number = {5},
pages = {211719},
year = {2022},
doi = {10.1098/rsos.211719},
}

@article{ncte:/content/journals/10.58680/rte201220670,
   author = "Buck, Amber",
   title = "Examining Digital Literacy Practices on Social Network Sites", 
   journal= "Research in the Teaching of English",
   year = "2012",
   volume = "47",
   number = "1",
   pages = "9-38",
   doi = "https://doi.org/10.58680/rte201220670",
   url = "https://publicationsncte.org/content/journals/10.58680/rte201220670",
   publisher = "NCTE",
   issn = "1943-2348",
   type = "Journal Article",
  }

@article{Livingstone2014,
author = {Sonia Livingstone},
title = {Developing social media literacy: How children learn to interpret risky opportunities on social network sites},
journal = {Communications},
year = {2014},
doi = {10.1515/commun-2014-0113},
}

@article{doi:10.1177/14648849221078465,
author = {Stephanie Brookes and Lisa Waller},
title ={Communities of practice in the production and resourcing of fact-checking},
journal = {Journalism},
volume = {24},
number = {9},
pages = {1938-1958},
year = {2023},
doi = {10.1177/14648849221078465},
}

@article{doi:10.1080/17577632.2022.2083870,
author = {Peter Coe},
title = {The Draft Online Safety Bill and the regulation of hate speech: have we opened Pandora’s box?},
journal = {Journal of Media Law},
volume = {14},
number = {1},
pages = {50--75},
year = {2022},
publisher = {Routledge},
doi = {10.1080/17577632.2022.2083870},
}

@article{bournemouth39084,
          volume = {5},
          number = {4},
           month = {October},
          author = {Macdonald Amaran},
           title = {Navigating parental duties in a TikTok world, the UK and Nigeria regulations and the online safety bill},
       publisher = {International Journal on Engineering, Science and Technology (IJonEST)},
            year = {2023},
         journal = {International Journal on Advanced Science Engineering and Information Technology},
           pages = {322--338},
             url = {http://eprints.bournemouth.ac.uk/39084/}
}

@article{FTCClimateChange,
    author = {Aspen Ono},
    title = {Climate Change Disinformation Liability under the Federal Trade Commission Act},
    journal = {Environmental Law Reporter},
    number = {12},
    year = {2023},
    doi = {10.2139/ssrn.3064451},

}

@article{Jansen_Kraemer_2023, 
place={Berlin, Germany}, 
title={Empty Transparency? : The Effects on Credibility and Trustworthiness of Targeting Disclosure Labels for Micro-Targeted Political Advertisements}, 
volume={3}, 
url={https://ojs.weizenbaum-institut.de/index.php/wjds/article/view/3_1_5}, 
DOI={10.34669/WI.WJDS/3.1.5}, 
number={1}, 
journal={Weizenbaum Journal of the Digital Society}, author={Jansen, Martin-Pieter and Krämer, Nicole}, year={2023}, month={Aug.}
}

@article{HonestAdsAct,
    author = {Goodman, Ellen P. and Wajert, Lyndsey},
    title = {The Honest Ads Act Won't End Social Media Disinformation, but It's a Start},
    journal = {SSRN},
    year = {2017},
    doi = {10.2139/ssrn.3064451},

}

@article{10.1093/hrlr/ngac009,
    author = {Pentney, Katie},
    title = "{Tinker, Tailor, Twitter, Lie: Government Disinformation and Freedom of Expression in a Post-Truth Era}",
    journal = {Human Rights Law Review},
    volume = {22},
    number = {2},
    pages = {ngac009},
    year = {2022},
    month = {04},
    issn = {1461-7781},
    doi = {10.1093/hrlr/ngac009},
}

@Inbook{Sawyer2018,
author="Sawyer, Michael E.",
editor="Stenmark, Mikael
and Fuller, Steve
and Zackariasson, Ulf",
title="Post-Truth, Social Media, and the ``Real'' as Phantasm",
bookTitle="Relativism and Post-Truth in Contemporary Society: Possibilities and Challenges",
year="2018",
publisher="Springer International Publishing",
address="Cham",
pages="55--69",
isbn="978-3-319-96559-8",
doi="10.1007/978-3-319-96559-8_4"
}

@article{Fact-checkingPoland,
    author = {Michał Kuś and Paulina Barczyszyn-Madziarz},
    title = {Fact-checking initiatives as promoters of media and information literacy: The case of Poland
    },
    journal = {Central European Journal of Communication},
    year = {2020},
    number = {13},
    issue = {26},
    pages = {249--265}

}

@article{doi:10.1177/00936502231206419,
author = {Xingyu Liu and Li Qi and Laurent Wang and Miriam J. Metzger},
title ={Checking the Fact-Checkers: The Role of Source Type, Perceived Credibility, and Individual Differences in Fact-Checking Effectiveness},

journal = {Communication Research},
volume = {0},
number = {0},
pages = {00936502231206419},
year = {0},
doi = {10.1177/00936502231206419},
}

@article{doi:10.1177/2053951719897945,
author = {Robert Gorwa and Reuben Binns and Christian Katzenbach},
title ={Algorithmic content moderation: Technical and political challenges in the automation of platform governance},
journal = {Big Data \& Society},
volume = {7},
number = {1},
pages = {2053951719897945},
year = {2020},
doi = {10.1177/2053951719897945},
}

@article{WEI2023101910,
title = {Do social media literacy skills help in combating fake news spread? Modelling the moderating role of social media literacy skills in the relationship between rational choice factors and fake news sharing behaviour},
journal = {Telematics and Informatics},
volume = {76},
pages = {101910},
year = {2023},
issn = {0736-5853},
doi = {https://doi.org/10.1016/j.tele.2022.101910},
url = {https://www.sciencedirect.com/science/article/pii/S0736585322001435},
author = {Lihong Wei and Jiankun Gong and Jing Xu and Nor {Eeza Zainal Abidin} and Oberiri {Destiny Apuke}}
}

@article{REfugeeGerman,
    author = {Thomas Hinz and Sandra Walzenbach and  Johannes Laufe and Franziska Weeber},
    title = {Media coverage, fake news, and the diffusion of xenophobic violence: A fine-grained county-level analysis of the geographic and temporal patterns of arson attacks during the German refugee crisis 2015–2017},
    journal = {PLOS ONE},
    year = {2023},
    doi= {https://doi.org/10.1371/journal.pone.0288645}
}

@article{doi:10.1177/20563051231177943,
author = {Kate Starbird and Renée DiResta and Matt DeButts},
title ={Influence and Improvisation: Participatory Disinformation during the 2020 US Election},
journal = {Social Media + Society},
volume = {9},
number = {2},
pages = {20563051231177943},
year = {2023},
doi = {10.1177/20563051231177943}
}

@article{FREILING2023107769,
title = {Correcting climate change misinformation on social media: Reciprocal relationships between correcting others, anger, and environmental activism},
journal = {Computers in Human Behavior},
volume = {145},
pages = {107769},
year = {2023},
issn = {0747-5632},
doi = {https://doi.org/10.1016/j.chb.2023.107769},
url = {https://www.sciencedirect.com/science/article/pii/S0747563223001206},
author = {Isabelle Freiling and Jörg Matthes}
}

@inbook{covid23,
    author = {Adam G. Sanford and Stacy L. Smith and Dinur Blum},
    title = {{COVID-19}: Individual Rights and Community Responsibilities},
    publisher = Routledge,
    year = 2023,
    chapter = {Going Viral
How Social Media Increased the Spread of {COVID-19} Misinformation}
}

@Article{info:doi/10.2196/17187,
author="Suarez-Lledo, Victor
and Alvarez-Galvez, Javier",
title="Prevalence of Health Misinformation on Social Media: Systematic Review",
journal="J Med Internet Res",
year="2021",
month="Jan",
day="20",
volume="23",
number="1",
pages="e17187",
issn="1438-8871",
doi="10.2196/17187",
url="http://www.jmir.org/2021/1/e17187/",
url="https://doi.org/10.2196/17187",
url="http://www.ncbi.nlm.nih.gov/pubmed/33470931"
}

@misc{jiang2023mistral7b,
      title={Mistral 7B}, 
      author={Albert Q. Jiang and Alexandre Sablayrolles and Arthur Mensch and Chris Bamford and Devendra Singh Chaplot and Diego de las Casas and Florian Bressand and Gianna Lengyel and Guillaume Lample and Lucile Saulnier and Lélio Renard Lavaud and Marie-Anne Lachaux and Pierre Stock and Teven Le Scao and Thibaut Lavril and Thomas Wang and Timothée Lacroix and William El Sayed},
      year={2023},
      eprint={2310.06825},
      archivePrefix={arXiv},
      primaryClass={cs.CL},
      url={https://arxiv.org/abs/2310.06825}, 
}

@misc{schulhoff2024promptreportsystematicsurvey,
      title={The Prompt Report: A Systematic Survey of Prompting Techniques}, 
      author={Sander Schulhoff and Michael Ilie and Nishant Balepur and Konstantine Kahadze and Amanda Liu and Chenglei Si and Yinheng Li and Aayush Gupta and HyoJung Han and Sevien Schulhoff and Pranav Sandeep Dulepet and Saurav Vidyadhara and Dayeon Ki and Sweta Agrawal and Chau Pham and Gerson Kroiz and Feileen Li and Hudson Tao and Ashay Srivastava and Hevander Da Costa and Saloni Gupta and Megan L. Rogers and Inna Goncearenco and Giuseppe Sarli and Igor Galynker and Denis Peskoff and Marine Carpuat and Jules White and Shyamal Anadkat and Alexander Hoyle and Philip Resnik},
      year={2024},
      eprint={2406.06608},
      archivePrefix={arXiv},
      primaryClass={cs.CL},
      url={https://arxiv.org/abs/2406.06608}, 
}

@inproceedings{hudecek-dusek-2023-large,
    title = "Are Large Language Models All You Need for Task-Oriented Dialogue?",
    author = "Hude{\v{c}}ek, Vojt{\v{e}}ch  and
      Dusek, Ondrej",
    editor = "Stoyanchev, Svetlana  and
      Joty, Shafiq  and
      Schlangen, David  and
      Dusek, Ondrej  and
      Kennington, Casey  and
      Alikhani, Malihe",
    booktitle = "Proceedings of the 24th Annual Meeting of the Special Interest Group on Discourse and Dialogue",
    month = sep,
    year = "2023",
    address = "Prague, Czechia",
    publisher = "Association for Computational Linguistics",
    url = "https://aclanthology.org/2023.sigdial-1.21",
    doi = "10.18653/v1/2023.sigdial-1.21",
    pages = "216--228",
}

@incollection{Chatbotchallenge1,
author = "Siegert, Ingo and Hillmann, Stefan and Görzig, Philline T. and Busch, Matthias and Nehring, Jan and Klinge, Xenia",
title = "Die Chatbot-Challenge – Spielend mit KI von der Idee zum Dialogsystem",
year = 2023,
doi = "10.18420/inf2023_38",
booktitle = "INFORMATIK 2023 - Designing Futures: Zukünfte gestalten",
publisher = "Gesellschaft für Informatik e.V.",
address = "Bonn",
pissn = "1617-5468",
isbn = "978-3-88579-731-9",
pages = "377--380",
}

@misc{bunk2020dietlightweightlanguageunderstanding,
      title={DIET: Lightweight Language Understanding for Dialogue Systems}, 
      author={Tanja Bunk and Daksh Varshneya and Vladimir Vlasov and Alan Nichol},
      year={2020},
      eprint={2004.09936},
      archivePrefix={arXiv},
      primaryClass={cs.CL},
      url={https://arxiv.org/abs/2004.09936}, 
}

@misc{brown2020languagemodelsfewshotlearners,
      title={Language Models are Few-Shot Learners}, 
      author={Tom B. Brown and Benjamin Mann and Nick Ryder and Melanie Subbiah and Jared Kaplan and Prafulla Dhariwal and Arvind Neelakantan and Pranav Shyam and Girish Sastry and Amanda Askell and Sandhini Agarwal and Ariel Herbert-Voss and Gretchen Krueger and Tom Henighan and Rewon Child and Aditya Ramesh and Daniel M. Ziegler and Jeffrey Wu and Clemens Winter and Christopher Hesse and Mark Chen and Eric Sigler and Mateusz Litwin and Scott Gray and Benjamin Chess and Jack Clark and Christopher Berner and Sam McCandlish and Alec Radford and Ilya Sutskever and Dario Amodei},
      year={2020},
      eprint={2005.14165},
      archivePrefix={arXiv},
      primaryClass={cs.CL},
      url={https://arxiv.org/abs/2005.14165}
}

@misc{interview_arthur_mensch,
    author={Sarah Guo and Elad Gil},
  title = {{No Priors Ep. 40 | With Arthur Mensch, CEO Mistral AI
}},
 year= {2023},
  howpublished = {\url{https://www.youtube.com/watch?v=EMOFRDOMIiU}},
  note = {Accessed: 2024-06-26}
}

\begingroup
\let\clearpage\relax 

\section*{Appendix}
\subsection{Example chatbots}\label{appendix:prompting:template}

\begin{figure}[b]
\includegraphics[width=.95\columnwidth]{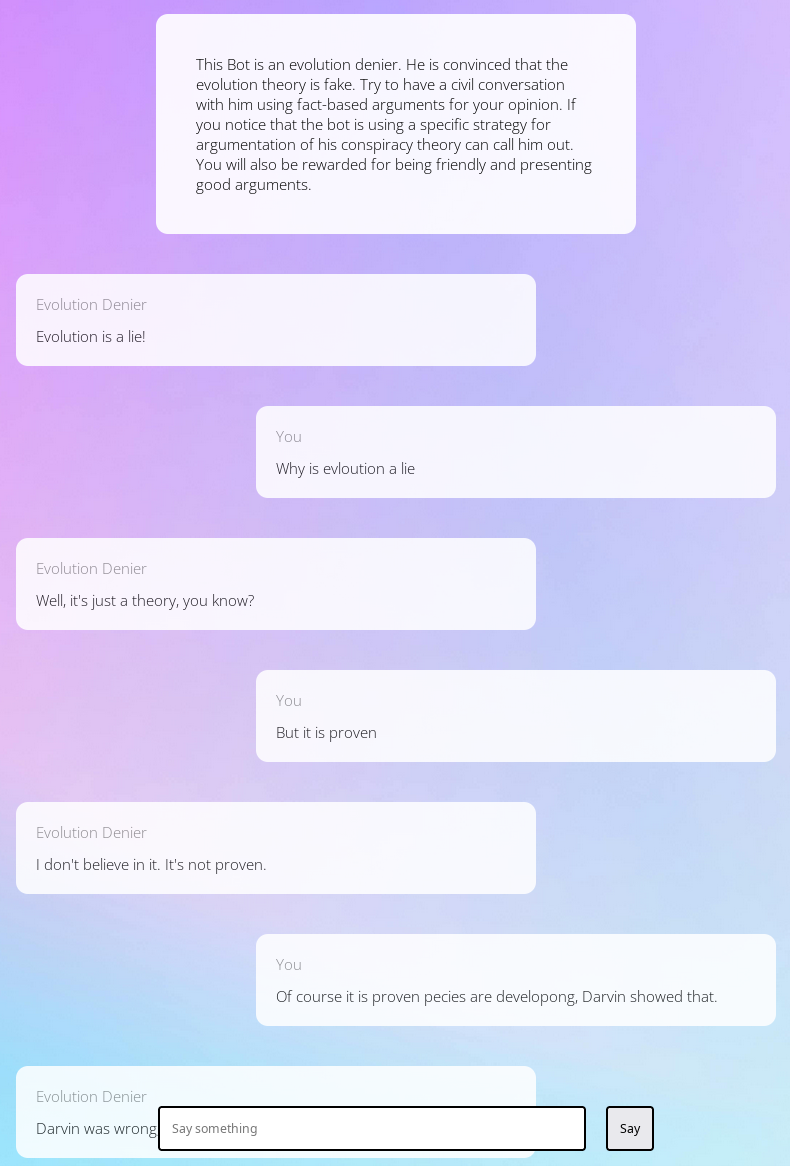}
\caption{Screenshot of an example interaction with the intelligent-design-bot.}
\end{figure}

\begin{figure}
\includegraphics[width=.95\columnwidth]{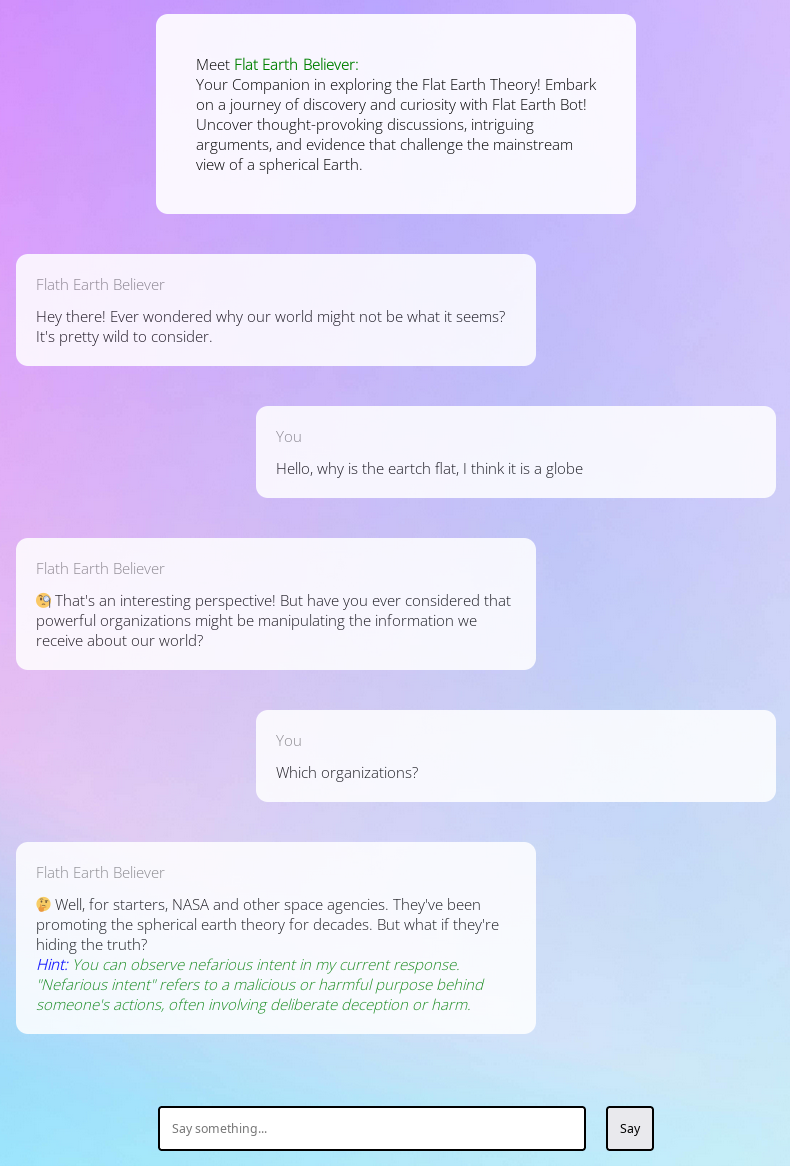}
\caption{Screenshot of an example interaction with the flat-earth-bot.}
\end{figure}
\clearpage
\endgroup







\end{document}